\documentclass[unsortedaddress,twocolumn,pre,amsmath,amssymb]{revtex4}
\usepackage[dvips]{graphicx}
\usepackage{dcolumn}
\usepackage{bm}
\begin{document}
\title{An information-based traffic control in a public conveyance system:\\ 
reduced clustering and enhanced efficiency} 
\author{Akiyasu Tomoeda and Katsuhiro Nishinari}
\affiliation{%
Department of Aeronautics and Astronautics,
Faculty of Engineering, University of Tokyo,
Hongo, Bunkyo-ku, Tokyo 113-8656, Japan.
}%
\author{Debashish Chowdhury}
\affiliation{%
Department of Physics, Indian Institute of Technology,
Kanpur 208016, India.
}%
\author{Andreas Schadschneider}%
\affiliation{%
Institut  f\"ur Theoretische  Physik, Universit\"at 
zu K\"oln D-50937 K\"oln, Germany
}%
\date{\today}
\begin{abstract}
A new public conveyance model applicable to buses and trains is proposed
 in this paper by using stochastic cellular automaton.
We have found the optimal density of vehicles, at which the average velocity
 becomes maximum, significantly
 depends on the number of stops and passengers behavior of getting
 on a vehicle at stops. The efficiency of the hail-and-ride system is also
 discussed by comparing the different behavior of passengers.
Moreover, we have found that a big cluster of vehicles is divided into small
 clusters, by incorporating information of the number of vehicles
 between successive stops.
\end{abstract}
\maketitle
\section{INTRODUCTION}

The totally asymmetric simple exclusion process \cite{sz,derrida,schuetz} is 
the simplest model of non-equilibrium systems of interacting self-driven
particles. Various extensions of this model have been reported in the 
last few years for capturing the essential features of the collective 
spatio-temporal organizations in wide varieties of systems, including 
those in vehicular traffic \cite{css,helbing,Schad,nagatanirev,cnss}. Traffic 
of buses and bicycles have also been modeled following similar approaches 
\cite{oloan,jiang}. A simple bus route model \cite{oloan} exhibits 
clustering of the buses along the route and the quantitative features 
of the coarsening of the clusters have strong similarities with 
coarsening phenomena in many other physical systems. Under normal 
circumstances, such clustering of buses is undesirable in any real 
bus route as the efficiency of the transport system is adversely 
affected by clustering. The main aim of this paper is to introduce a  
traffic control system into the bus route model in such a way that 
helps in suppressing this tendency of clustering of the buses. This 
new model exhibits a competition between the two opposing tendencies 
of clustering and de-clustering which is interesting from the point 
of view of fundamental physical principles. However, the model may 
also find application in developing adaptive traffic control systems 
for public conveyance systems.

In some of earlier bus-route models, movement of the buses was monitored 
on coarse time intervals so that the details of the dynamics of the 
buses in between two successive bus stops was not described explicitly.  
Instead, the movement of the bus from one stop to the next was captured 
only through probabilities of hopping from one stop to the next; hopping 
takes place with the lower probability if passengers are waiting at the 
approaching bus stop \cite{oloan}. An alternative interpretation of the 
model is as follows: the passengers could board the bus whenever and 
wherever they stopped a bus by raising their hand, this is called the 
{\it hail-and-ride} system.

Several possible extensions of the bus route model have been reported 
in the past \cite{cd,nagatani,Chi}. For example, in \cite{cd},
in order to elucidate the connection between the bus route model with
parallel updating and the Nagel-Schreckenberg model, two alternative
extensions of the latter model with space-/time-dependent hopping 
rates are proposed. If a bus does not stop at a bus stop, the 
waiting passengers have to wait further for the next bus; such 
scenarios were captured in one of the earlier bus route models 
\cite{nagatani}, using modified car-following model. In \cite{Chi}, 
the bus capacity, as well as the number of passengers getting on and 
off at each stop, were introduced to make the model more realistic. 
Interestingly, it has been claimed that the distribution of the time 
gaps between the arrival of successive buses is described well by the 
Gaussian Unitary Ensemble of random matrices \cite{Mex}.

In this paper, by extending the model in \cite{oloan}, we suggest a new 
public conveyance model (PCM). Although we refer to each of the public 
vehicles in this model as a ``bus'', the model is equally applicable 
to train traffic on a given route. In this PCM we can set up arbitrary 
number of bus stops on the given route. The {\it hail-and-ride} system 
turns out to be a special case of the general PCM. Moreover, in 
the PCM the duration of the halt of a bus at any arbitrary bus stop 
depends on the number of waiting passengers. As we shall demonstrate 
in this paper, the delay in the departure of the buses from crowded 
bus stops leads to the tendency of the buses to cluster on the route. 
Furthermore, in the PCM, we also introduce a traffic control system that 
exploits the information on the number of buses in the ``segments'' 
in  between successive bus stops; this traffic control system helps 
in reducing the undesirable tendency of clustering by dispersing the 
buses more or less uniformly along the route.

In this study we introduce two different quantitative measures of 
the efficiency of the bus transport system, and calculate these 
quantities, both numerically and analytically, to determine the 
conditions under which the system would operate optimally.

This paper is organized as follows, in Sec.~$2$ PCM is introduced
 and we show several simulation results in Sec.~$3$.
The average speed and the number of waiting
passengers are studied by mean field
analysis in Sec.~$4$, and conclusions are given in Sec.~$5$.

\section{A STOCHASTIC CA MODEL FOR PUBLIC CONVEYANCE}

In this section, we explain the PCM in detail. For the sake of simplicity, 
we impose periodic boundary conditions. Let us imagine that the road is 
partitioned into $L$ identical cells such that each cell can accommodate 
at most one bus at a time. Moreover, a total of $S$ ($0\le S \le L$) 
{\it equispaced} cells are identified in the beginning as bus stops. Note 
that, the special case $S=L$ corresponds to the {\it hail-and-ride} system.
At any given time step, a passenger arrives with probability $f$ to the
system. Here, we assume that a given passenger is equally likely to
arrive at any one of the bus stops with a probability $1/S$. Thus, the
average number of passengers that arrive at each bus stop per unit time
is given by $f/S$. In contrast to this model, in ref.~\cite{cgns,kjnsc}
the passengers were assumed to arrive with probability $f$ at all the
bus stops in every time step.

\begin{figure}[h]
\begin{center}
\includegraphics[scale=0.6]{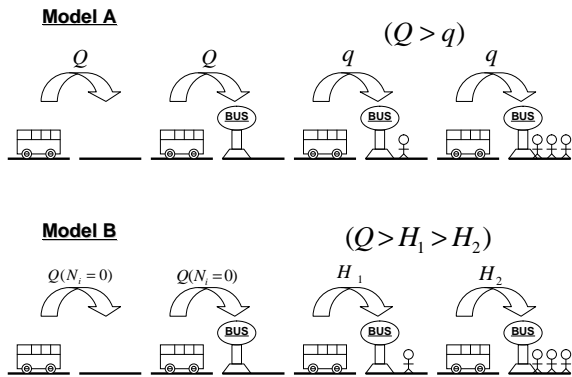}
\caption{Schematic illustration of the PCM. In the model A, the hopping
 probability to the bus stop does not depend on the number of waiting
 passengers. In contrast, in the model B the hopping probability to the
 bus stop depends on the number of waiting passengers. Thus if the
 waiting passengers increase, the hopping probability to the bus stop is
 decreased.}
\label{modelAB}
\end{center}
\end{figure}

The model A corresponds to those situations where, because of 
sufficiently large number of broad doors, the time interval during 
which the doors of the bus remain open after halting at a stop, is 
independent of the size of waiting crowd of passengers. In contrast, 
the model B captures those situations where a bus has to halt 
for a longer period to pick up a larger crowd of waiting passengers. 

The symbol $H$ is used to denote the hopping probability of a bus
entering into a cell that has been designated as a bus stop. We consider 
two different forms of $H$ in the two versions of our model which are 
named as model A and model B. In the model A we assume the form 
\begin{equation}
 H=\left\{\begin{array}{cl}
Q & {\rm \,\,\,\,\,\, no \,\, waiting \,\, passengers }\\
q & {\rm \,\,\,\,\,\, waiting \,\, passengers\,\, exist }
\end{array}
   \right.
\label{ant}
\end{equation}
where both $Q$ and $q$ ($Q > q$) are constants independent of the number 
of waiting passengers. The form (\ref{ant}) was used in the original 
formulation of the bus route model by O'Loan et al.\ \cite{oloan}.

In contrast to most of all the earlier bus route models, we assume 
in the model B that the maximum number of passengers that can get 
into one bus at a bus stop is $N_{\rm max}$. Suppose, $N_i$ denotes 
the number of passengers waiting at the bus stop $i$ $(i=1,\cdots,S)$ 
at the instant of time when a bus arrives there. In contrast to the 
form (\ref{ant}) for $H$ in model A, we assume in model B the form 
\begin{equation}
 H=\frac{Q}{\min(N_i,N_{\rm max})+1}
\label{wait}
\end{equation}
where $\min(N_i,N_{\rm max})$ is the number of passengers who can get 
into a bus which arrives at the bus stop $i$ at the instant of time 
when the number of passengers waiting there is $N_i$. The form 
(\ref{wait}) is motivated by the common expectation that the time 
needed for the passengers boarding a bus is proportional to their 
number. FIG.~$\ref{modelAB}$ depicts the hopping probabilities in
the two models A and B schematically. 

The hopping probability of a bus to the cells that are not designated 
as bus stops is $Q$; this is already captured by the expressions 
(\ref{ant}) and (\ref{wait}) since no passenger ever waits at those 
locations.

In principle, the hopping probability $H$ for a real bus would depend 
also on the number of passengers who get off at the bus stop; in the 
extreme situations where no passenger waits at a bus stop the hopping 
probability $H$ would be solely decided by the disembarking passengers. 
However, in order to keep the model theoretically simple and tractable,
we ignore the latter situation and assume that passengers get off only 
at those stops where waiting passengers get into the bus and that the 
time taken by the waiting passengers to get into the bus is always 
adequate for the disembarking passengers to get off the bus.

Note that $N_{\rm max}$ is the {\it maximum boarding capacity} at each bus 
stop rather than the {\it maximum carrying capacity} of each bus. 
The PCM model reported here can be easily extended to incorporate an 
additional dynamical variable associated with each bus to account for 
the instantaneous number of passengers in it. But, for the sake of 
simplicity, such an extension of the model is not reported here.
Instead, in the simple version of the PCM model reported here, $N_{\rm max}$ 
can be interpreted as the maximum carrying capacity of each bus if we 
assume that all of the passengers on the bus get off whenever it stops.


The model is updated according to the following rules. In step
$2-4$, these rules are applied in {\it parallel} to all
buses and passengers, respectively:
\begin{enumerate}
 \item {\it Arrival of a passenger}\\
A bus stop $i$ ($i=1,\cdots,S$) is picked up randomly, with probability 
$1/S$, and then the corresponding number of waiting passengers in 
increased by unity, i.e. $N_i$ $\rightarrow$ $N_i+1$, with probability 
$f$ to account for the arrival of a passenger at the selected bus stop.
 \item {\it Bus motion}\\
If the cell in front of a bus is not occupied by another bus,
each bus hops to the next cell with the probability $H$.
Specifically, if passengers do not exist in the next cell in both
       model A and model B hopping probability equals to $Q$ because
       $N_i$ equals to 0. Else, if passengers exist in the next cell, 
       the hopping probability equals to $q$ in the model A, whereas  
       in the model B the corresponding hopping probability equals to 
       $Q/(\min(N_i,N_{\rm max})+1)$. Note that, when a bus is 
       loaded with passengers to its maximum boarding capacity
       $N_{\rm max}$, the hopping probability in the model B equals to 
       $Q/(N_{\rm max}+1)$, the smallest allowed hopping probability. 
 \item {\it Boarding a bus}\\
When a bus arrives at the $i$-th ($i=1,\cdots,S$) bus stop cell, the 
corresponding number $N_i$ of waiting passengers is updated to 
$\max(N_i-N_{\rm max},0)$ to account for the passengers boarding the bus. 
Once the door is closed, no more waiting passenger can get into the bus 
at the same bus stop although the bus may remain stranded at the same 
stop for a longer period of time either because of the unavailability 
of the next bus stop or because of the traffic control rule explained 
next.
 \item {\it Bus information update}\\
Every bus stop has information $I_j$ ($j=1,\cdots,S$) which is the 
number of buses in the segment of the route between the stop $j$ and 
the next stop $j+1$ at that instant of time. This information is 
updated at each time steps. When one bus leaves the $j$-th bus stop, 
$I_j$ is increased to $I_j+1$. On the other hand, when a bus leaves 
$(j+1)$-th bus stop, $I_j$ is reduced to $I_j-1$. The desirable value 
of $I_j$ is $I_0 = m/S$, where $m$ is the total number of buses,
for all $j$ so that buses are not clustered 
in any segment of the route. We implement a traffic control rule 
based on the information $I_j$: a bus remains stranded at a stop $j$ 
as long as $I_j$ exceeds $I_0$. 
\end{enumerate}

We use the average speed $\langle V \rangle$ of the buses and the
number of the waiting passengers $\langle N \rangle$ at a bus stop
as two quantitative measures of the efficiency of the public conveyance 
system under consideration; a higher $\langle V \rangle$ and smaller 
$\langle N \rangle$ correspond to an efficient transportation system.

\section{COMPUTER SIMULATIONS OF PCM}

In the simulations we set $L=500, Q=0.9, q=0.5$ and $N_{\rm max}=60$.
The main parameters of this model, which we varied, are the number of 
buses ($m$), the number of bus stops ($S$) and the probability ($f$) 
of arrival of passengers. The number density of buses is defined by 
$\rho=m/L$.

\begin{figure}[h]
\begin{center}
\includegraphics[scale=0.8]{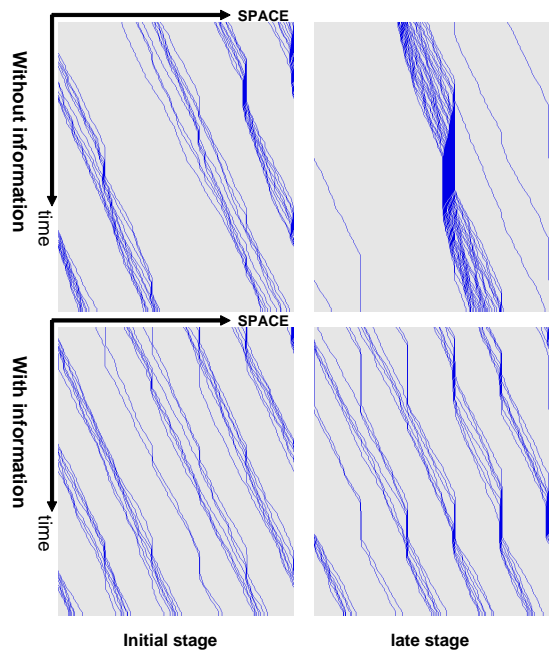}
\caption{Space-time plots in the model B for the parameter values 
$f=0.6, S=5, m=30$. The upper two figures correspond to the case 
where no traffic control system based on the information $\{I\}$ 
is operational. The upper left figure corresponds to the initial 
stage (from $t=1000$ to $t=1500$) whereas the upper right plot 
corresponds to the late stages (from $t=4000$ to $t=4500$). The 
lower figures correspond to the case where the information ($\{I\}$) 
based bus-traffic control system is operational (left figure 
shows data from $t=1000$ to $t=1500$ while the right figure 
corresponds to $t=4000$ to $t=4500$). Clearly, information-based 
traffic control system disperses the buses which, in the absence 
of this control system, would have a tendency to cluster.
}
\label{spatemp}
\end{center}
\end{figure}

Typical space-time plots of the model B are given in FIG.~\ref{spatemp}.
If no information-based traffic control system exits, the buses have a 
tendency to cluster; this phenomenon is very simular to that observed 
in the ant-trail model \cite{cgns,kjnsc}. However, implementation of 
the information-based traffic control system restricts the size of such 
clusters to a maximum of $I_0$ buses in a segment of the route in between 
two successive bus stops. We study the effects of this control system 
below by comparing the characteristics of two traffic systems one of 
which includes the information-based control system while the other 
does not.

\subsection{PCM without information-based traffic control} 
\begin{figure}[h]
\begin{center}
\includegraphics[scale=0.52]{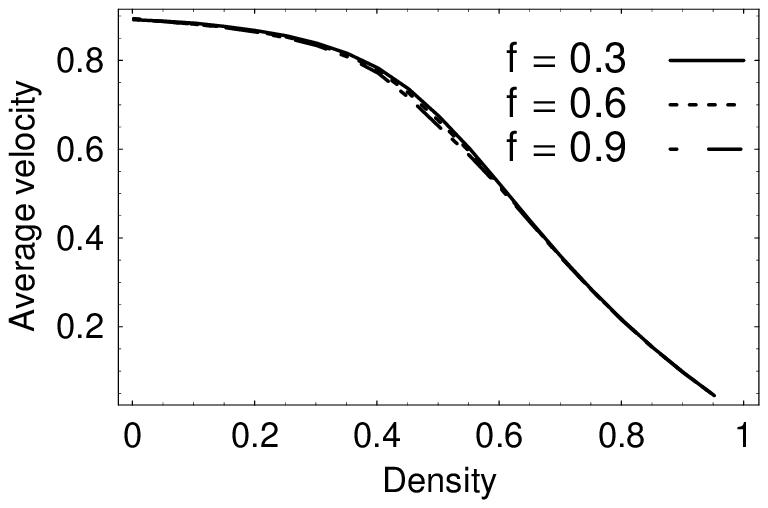}
\includegraphics[scale=0.52]{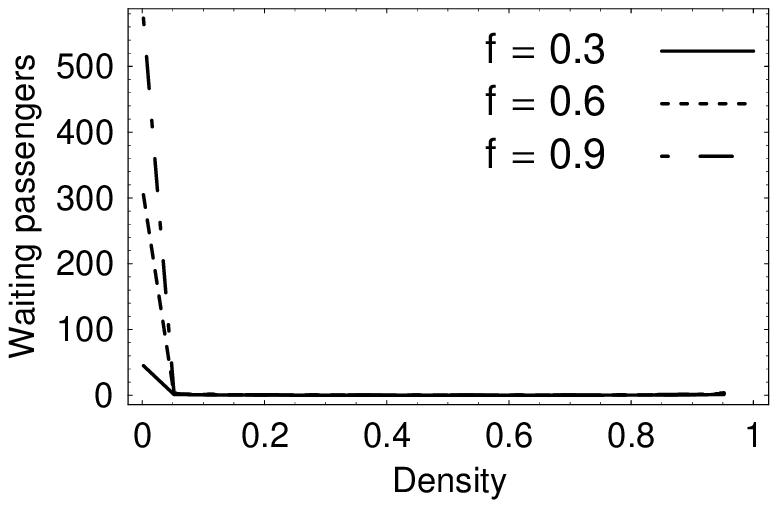}
\caption{The average speed and the average number of waiting passengers
 in the model A are plotted against the density for the parameters $S=5$
 and $f=0.3$, 0.6 and 0.9.}
\label{S=5_Rulefalse_noinfo}
\end{center}
\end{figure}

\begin{figure}[h]
\begin{center}
\includegraphics[scale=0.52]{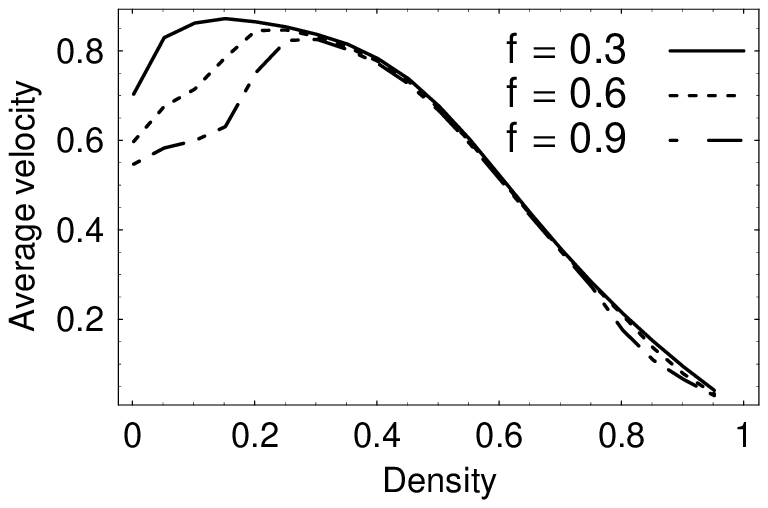}
\includegraphics[scale=0.52]{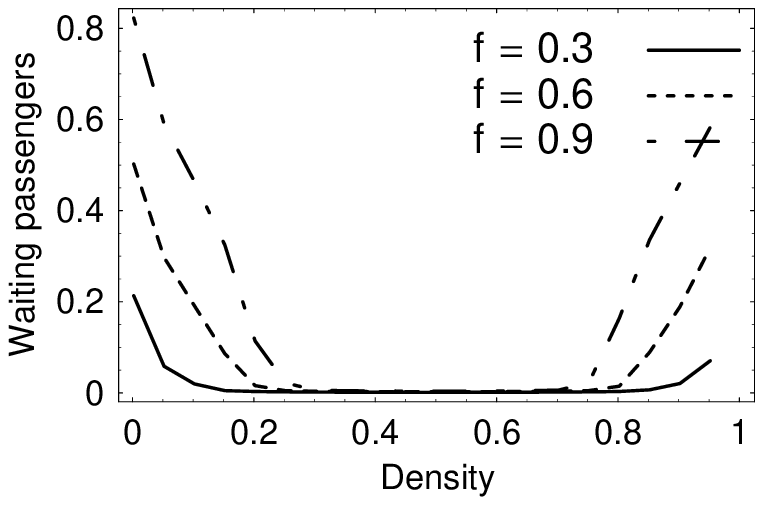}
\caption{The plot of $\langle V \rangle$ and $\langle N \rangle$ of the
 model A for $S=500(=L)$ and $f=0.3$, 0.6 and 0.9.}
\label{S=500_Rulefalse_noinfo}
\end{center}
\end{figure}

\begin{figure}[h]
\begin{center}
\includegraphics[scale=0.52]{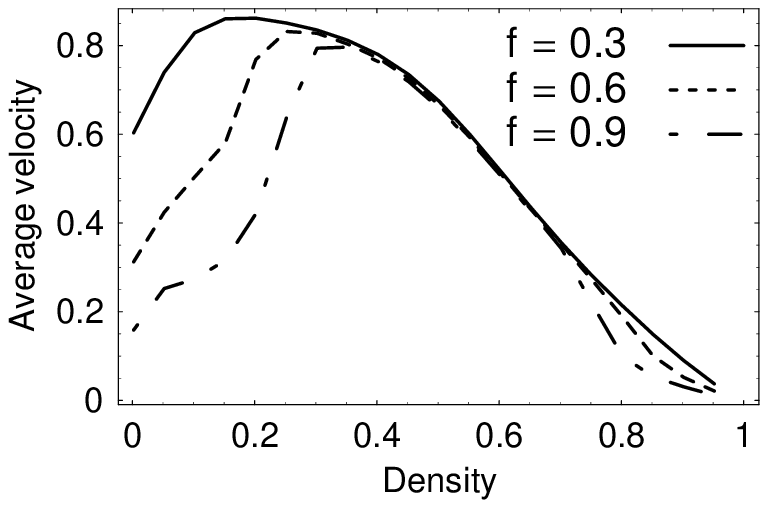}
\includegraphics[scale=0.5]{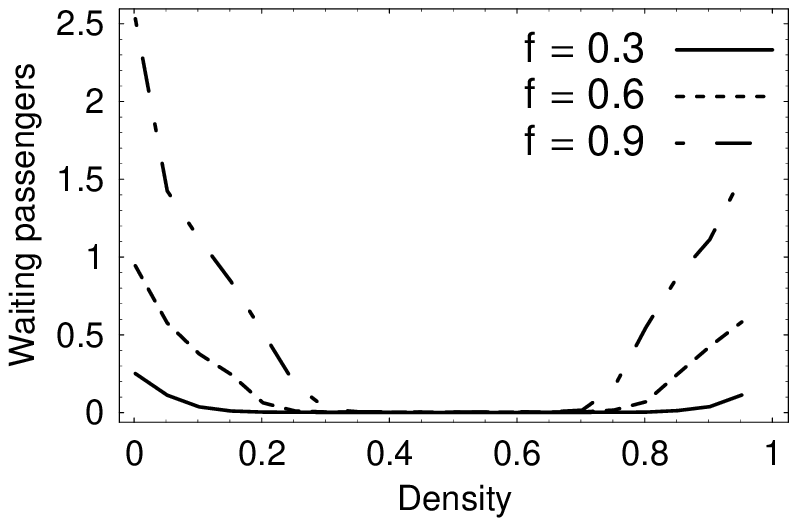}
\caption{The plot of $\langle V \rangle$ and $\langle N \rangle$ of the
 model B for $S=500(=L)$ and $f=0.3$, 0.6 and 0.9.}
\label{S=500_Ruletrue_noinfo}
\end{center}
\end{figure}

In the FIG.~\ref{S=5_Rulefalse_noinfo} - FIG.~\ref{S=500_Ruletrue_noinfo}, 
we plot $\langle V \rangle$ and $\langle N \rangle$ against the density
of buses for several different values of $f$.
Note that, the FIG.~\ref{S=500_Rulefalse_noinfo} and
FIG.~\ref{S=500_Ruletrue_noinfo} corresponds to the hail-and-ride system
for models A and B, respectively.

\begin{figure}[h]
\begin{center}
\includegraphics[scale=0.52]{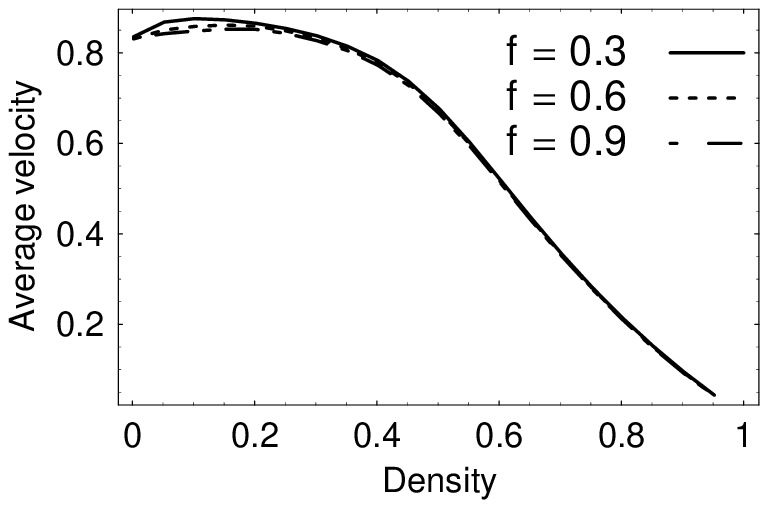}
\includegraphics[scale=0.52]{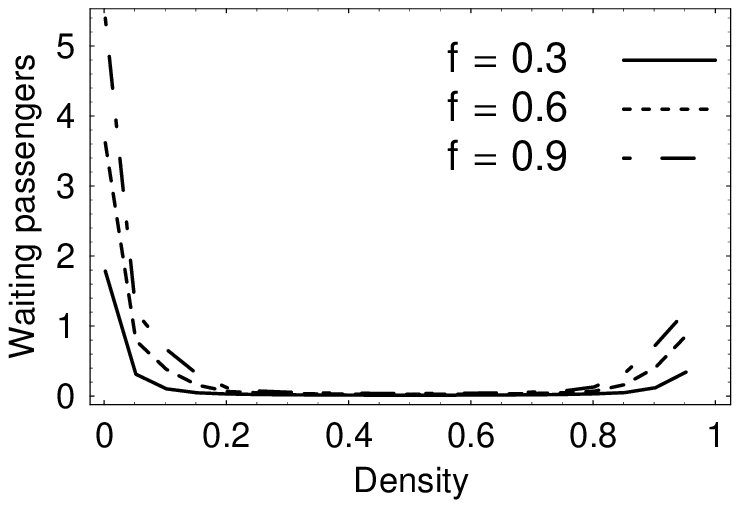}
\caption{The plot of $\langle V \rangle$ and $\langle N \rangle$ of the
 model A for $S=50$ and $f=0.3$, 0.6 and 0.9.}
\label{S=50_Rulefalse_noinfo}
\end{center}
\end{figure}

\begin{figure}[h]
\begin{center}
\includegraphics[scale=0.52]{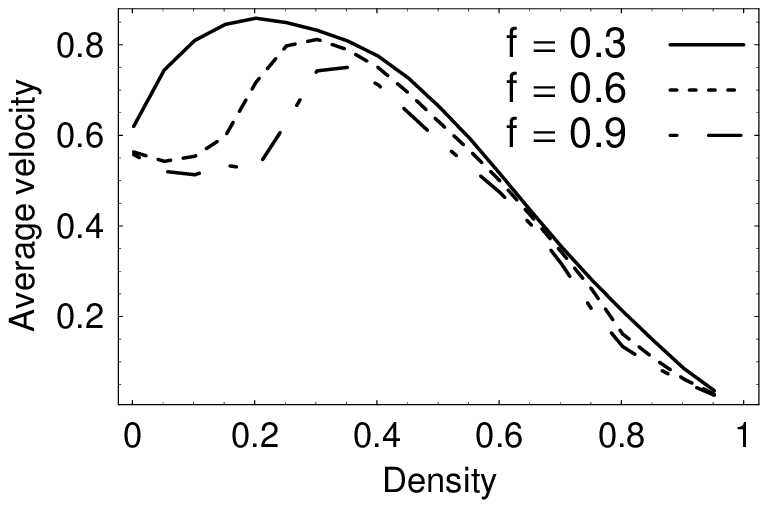}
\includegraphics[scale=0.5]{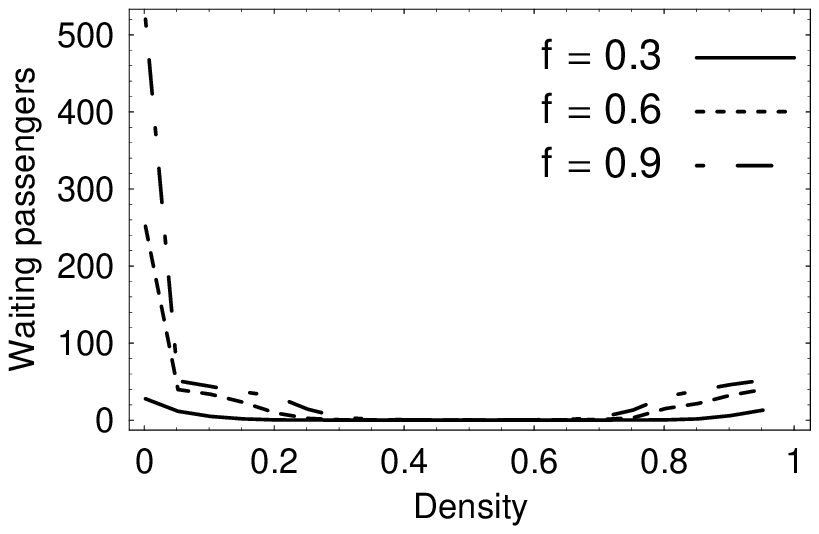}
\caption{The plot of $\langle V \rangle$ and $\langle N \rangle$ of the
 model B for $S=5$ and $f=0.3$, 0.6 and 0.9.}
\label{S=5_Ruletrue_noinfo}
\end{center}
\end{figure}

\begin{figure}[h]
\begin{center}
\includegraphics[scale=0.52]{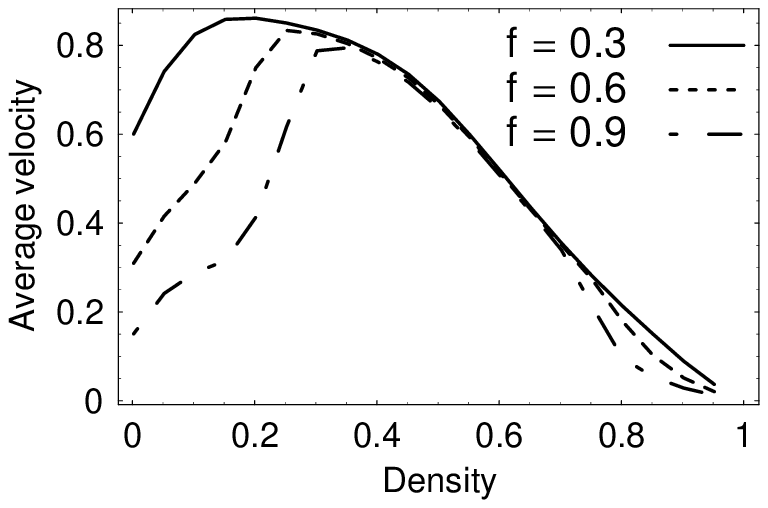}
\includegraphics[scale=0.5]{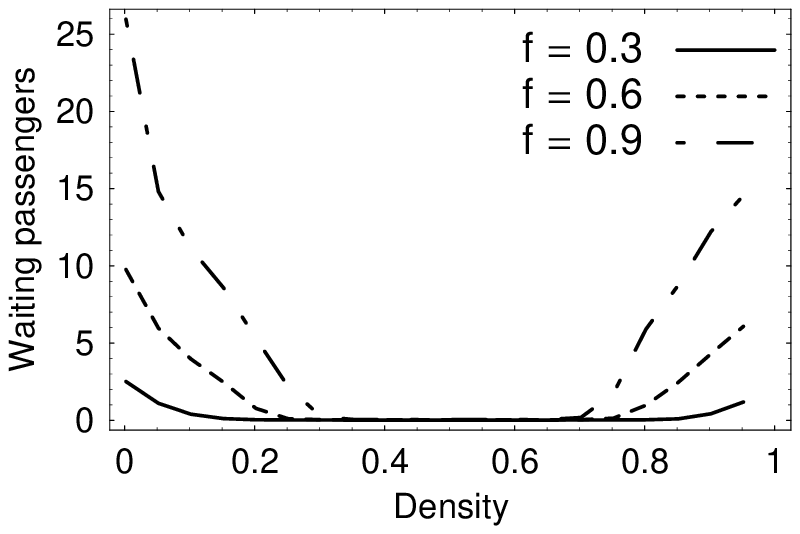}
\caption{The plot of $\langle V \rangle$ and $\langle N \rangle$ of the
 model B for $S=50$ and $f=0.3$, 0.6 and 0.9.}
\label{S=50_Ruletrue_noinfo}
\end{center}
\end{figure}

\begin{figure}[h]
\begin{center}
\includegraphics[scale=0.9]{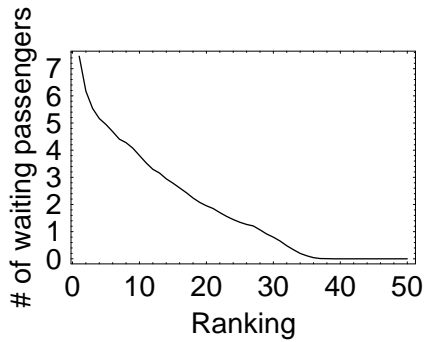}
\caption{The distribution of waiting passengers is plotted against
 all bus stops for the parameters $f=0.6$, $B=50$, $S=50$.
The horizontal line means the ranking, where we arrange the bus stops
 according to the descending order of $\langle N \rangle$.
}
\label{zip}
\end{center}
\end{figure}

\begin{figure}[h]
\begin{center}
\includegraphics[scale=0.52]{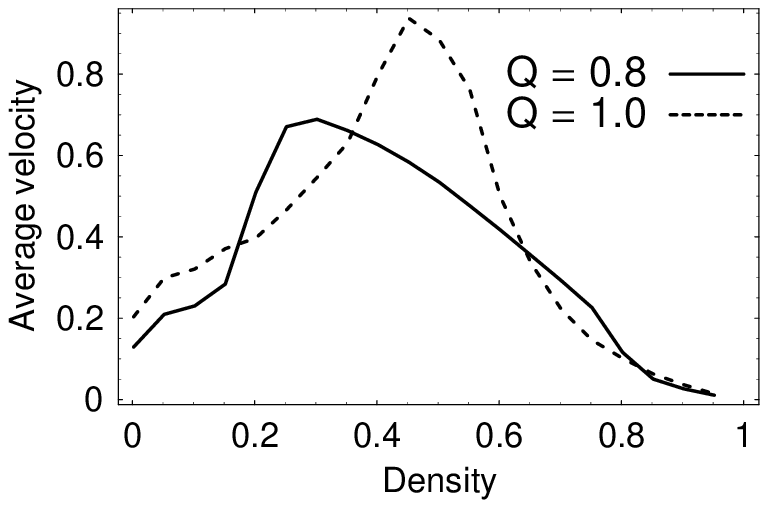}
\includegraphics[scale=0.52]{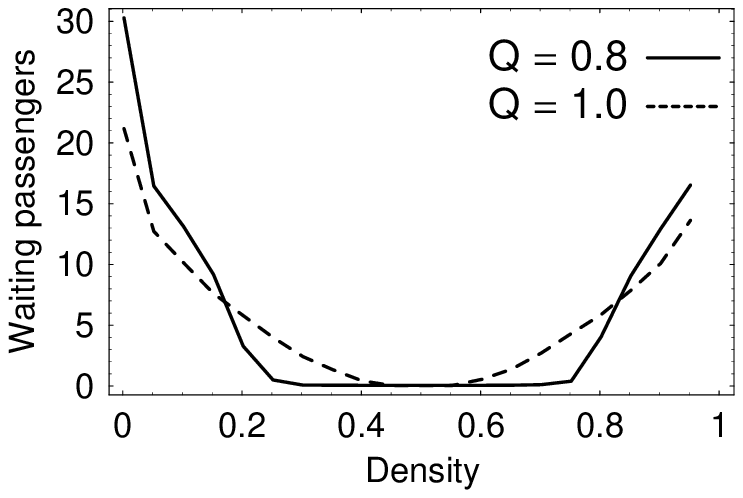}
\caption{The average speed and the average number of waiting passengers 
in the model B are plotted against the density for the parameters 
$f=0.9, S=50$; the hopping parameters are $Q=0.8$ and $Q=1.0$.
}
\label{FD4}
\end{center}
\end{figure}

\begin{figure}[h]
\begin{center}
\includegraphics[scale=0.8]{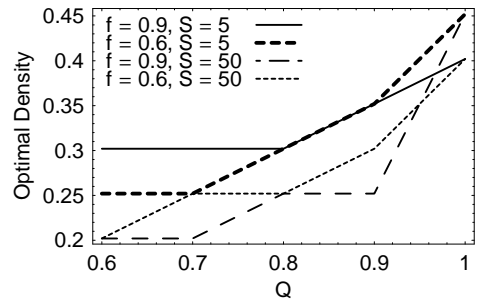}
\caption{The optimal density of buses in the model B is plotted 
against $Q$.  The parameters are
$f=0.9, S=5$ (normal line),$f=0.6, S=5$ (finer broken line),
$f=0.9, S=50$ (bold broken line),
$f=0.6, S=50$ (longer broken line).
}
\label{opt}
\end{center}
\end{figure}

\begin{figure}[h]
\begin{center}
\includegraphics[scale=0.7]{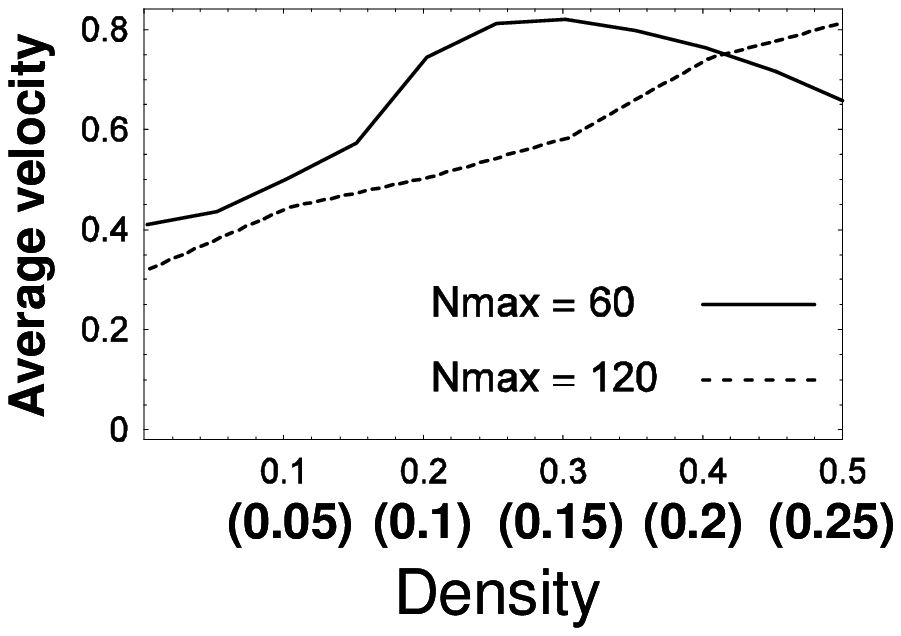}
\includegraphics[scale=0.7]{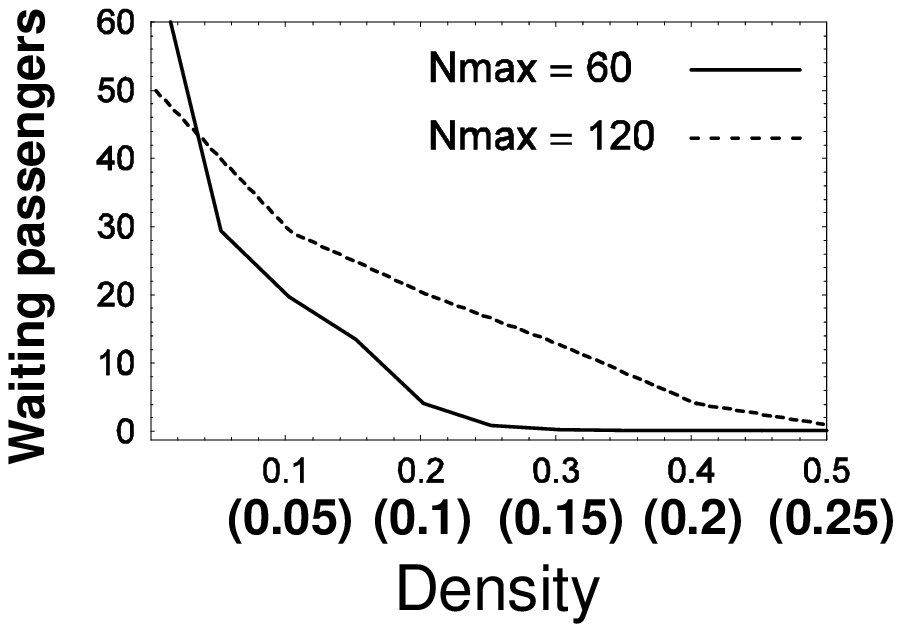}
\includegraphics[scale=0.7]{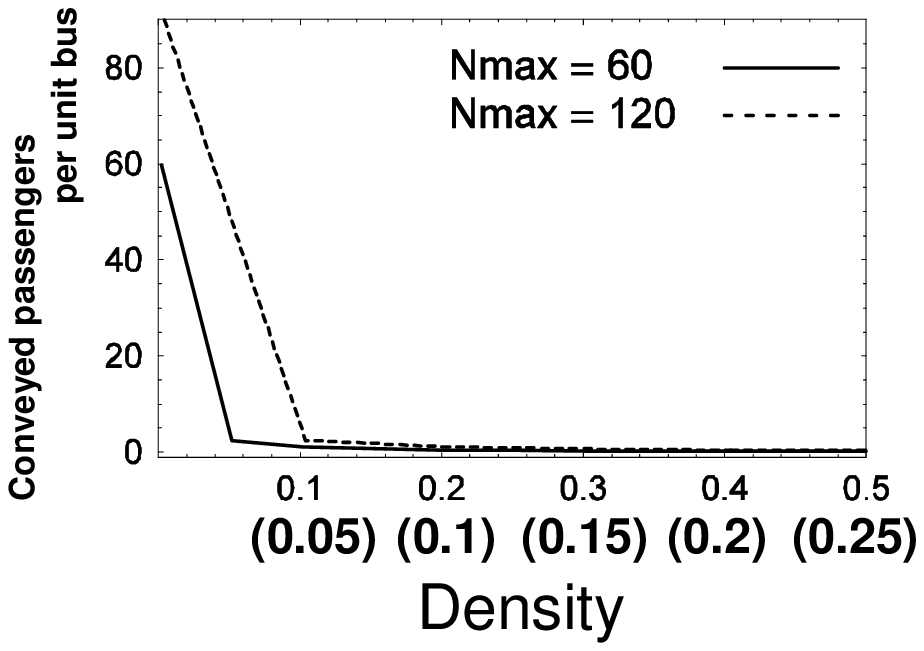}
\caption{Comparison between the case of bus capacity $60$ with
 bus capacity $120$.
The parameters are $Q=0.9$, $S=10$, $f=0.6$ in the model B without information.
The top figure shows the average velocity, the center figure shows
 waiting passengers and the bottom figure shows the number of conveyed
 passengers per unit bus, i.e. this number is calculated by (total number of
 on-boarding passengers on all buses)/(the number of buses),
against the bus density up to $0.5$. 
In each figure, the horizontal axis shows the density; the numbers 
without parentheses denote the number densities in the case $N_{\rm max}=60$, 
whereas the numbers in the parentheses denote the number densities in 
the case $N_{\rm max}=120$.}
\label{hiki}
\end{center}
\end{figure}

These figures demonstrate that the average speed $\langle V \rangle$, 
which is a measure of the efficiency of the bus traffic system, 
exhibits a {\it maximum} at around $\rho=0.2\sim0.3$ especially in the model
B (comparing FIG.~\ref{S=5_Rulefalse_noinfo} with
FIG.~\ref{S=5_Ruletrue_noinfo}, it shows the model
B (FIG.~\ref{S=5_Ruletrue_noinfo}) reflects the bus bunching more clearly
than the model A (FIG.~\ref{S=5_Rulefalse_noinfo}) especially at large f
and small $\rho$).
The average number of waiting passengers $\langle N \rangle$, whose
inverse is another measure of the efficiency of the bus traffic system,
is vanishingly small in the region $0.3 < \rho < 0.7$; $\langle N
\rangle$ increases with decreasing (increasing) $\rho$ in the regime
$\rho < 0.3$ ($\rho > 0.7$). 

The average velocity of the model A becomes smaller as S increases in
the low density region (see
FIG.~\ref{S=5_Rulefalse_noinfo}, FIG.~\ref{S=50_Rulefalse_noinfo} and
FIG.~\ref{S=500_Rulefalse_noinfo}).
In contrast, in the model B (FIG.~\ref{S=50_Ruletrue_noinfo} and
FIG.~\ref{S=500_Ruletrue_noinfo})
we observe that there is no significant difference in the average
velocity.
Note that the number of waiting passengers is calculated by (total
waiting passengers)/(number of bus stops). The total number of waiting 
passengers in this system is almost the same under the case $S=50$ and
hail-and-ride system $S=L$ in both models.
When the parameter $S$ is small (comparing FIG.~\ref{S=5_Rulefalse_noinfo}
and FIG.~\ref{S=5_Ruletrue_noinfo}), in the model B the waiting
passengers are larger and the average velocity is smaller than in the
model A, since the effect of the delay in getting on a bus is taken into
account. In the model B (comparing FIG.~\ref{S=5_Ruletrue_noinfo},
FIG.~\ref{S=50_Ruletrue_noinfo} and FIG.~\ref{S=500_Ruletrue_noinfo}),
the case $S=50$ is more efficient than $S=5$, i.e. the system is likely
to become more efficient, as $S$ increases. However, we do not find any 
significant variation between $S=50$ and $S=500$. When $S$ is small, 
the system becomes more efficient by increasing the number of bus
stops. 
If the number of bus stops increase beyond $50$, then there is little
further variation of the efficiency as $S$ is increased up to the maximum 
value $500$.

From FIG.~\ref{zip}, the distribution of $\langle N \rangle$ over all the
bus stops in the system is shown.
We see that the distribution does not show the Zipf's law, which is
sometimes seen in natural and social phenomena; frequency of used
words \cite{word}, population of a city \cite{population}, the number of
the access to a web site \cite{web}, and intervals between successive
transit times of the cars of traffic flow \cite{musha}.

Next, we investigate the optimal density of buses at which the average
velocity becomes maximum. The optimal density depends on $Q$ and is 
$\rho=0.3$ for $Q=0.8$ (FIG.~\ref{FD4}, see also FIG.~\ref{opt}).
In FIG.~\ref{FD4}, it is shown that the density corresponding to the 
maximum velocity shifts to higher values as $Q$ becomes larger.
FIG.~\ref{opt} shows the optimal density of buses in the model B 
without information-based control system. From this figure, we find 
that the optimal density, for case $S=50$, is smaller than that for 
$S=5$. Moreover, for given $S$, the optimal density decreases with 
decreasing $f$. However, for both $S=5$ and $S=50$, the optimal 
density corresponding to $Q=1.0$ is higher for $f=0.6$ than that for 
$f=0.9$. 

What is more effective way of increasing the efficiency of the public 
conveyance system on a given route by increasing the number of buses    
without increasing the carrying capacity of each bus, or by increasing 
the carrying capacity of each bus without recruiting more buses? Or, 
are these two prescriptions for enhancing efficiency of the public 
conveyance system equally effective? In order to address these questions, 
we make a comparative study of two situations on the same route: for
example, in the first situation the number of buses is $10$ and each has
a capacity of $60$, whereas in the second the number of buses is $5$ and
each has a capacity of $120$. Note that the total carrying capacity 
of all the buses together is $600$ ($60\times 10$ and $120\times 5$ in 
the two situations), i.e., same in both the situations. But, the number 
density of the buses in the second situation is just half of that in 
the first as the length of the bus route is same in both the situations. 
In FIG.~$\ref{hiki}$, the results for these two cases are plotted; the 
different scales of density used along the $X$-axis arises from the 
differences in the number densities mentioned above. 

From FIG.~$\ref{hiki}$, we conclude that, at sufficiently low
densities, the average velocity is higher for $N_{\rm max}=60$ compared to
those for $N_{\rm max}=120$. But, in the same regime of the number density
of buses, larger number of passengers wait at bus stops when the bus
capacity is smaller. Thus, in the region $\rho<0.05$, system
administrators face a dilemma: if they give priority to the average
velocity and decide to choose buses with $N_{\rm max}=60$, the number of
passengers waiting at the bus stops increases. On the other hand if they
decide to make the passengers happy by reducing their waiting time at
the bus stops and, therefore, choose buses with $N_{\rm max}=120$, the
travel time of the passengers after boarding a bus becomes longer.

However, at densities $\rho>0.05$, the system administrators can satisfy 
both the criteria, namely, fewer waiting passengers and shorter travel 
times, by one single choice. In this region of density, the public 
conveyance system with $N_{\rm max}=60$ is more efficient than that with 
$N_{\rm max}=120$ because the average velocity is higher and the number of 
waiting passengers is smaller for $N_{\rm max} = 60$ than for $N_{\rm max}=120$.
Thus, in this regime of bus density, efficiency of the system is enhanced 
by reducing the capacity of individual buses and increasing their number 
on the same bus route. 

\begin{figure}[h]
\begin{center}
\includegraphics[scale=0.52]{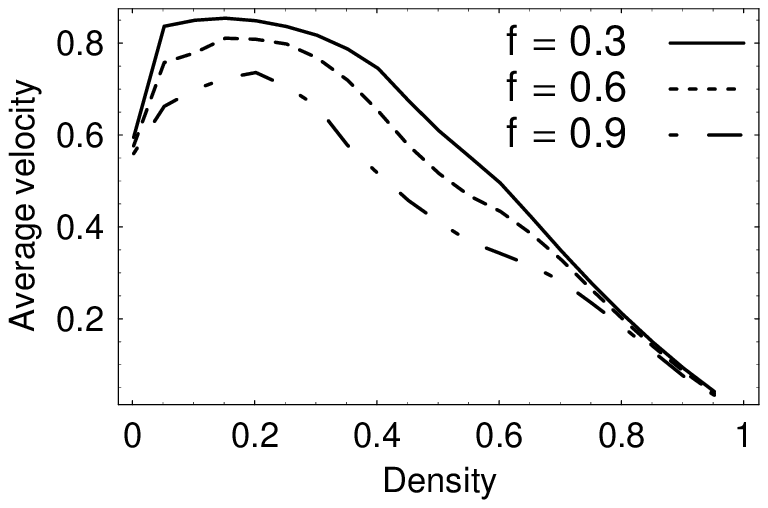}
\includegraphics[scale=0.52]{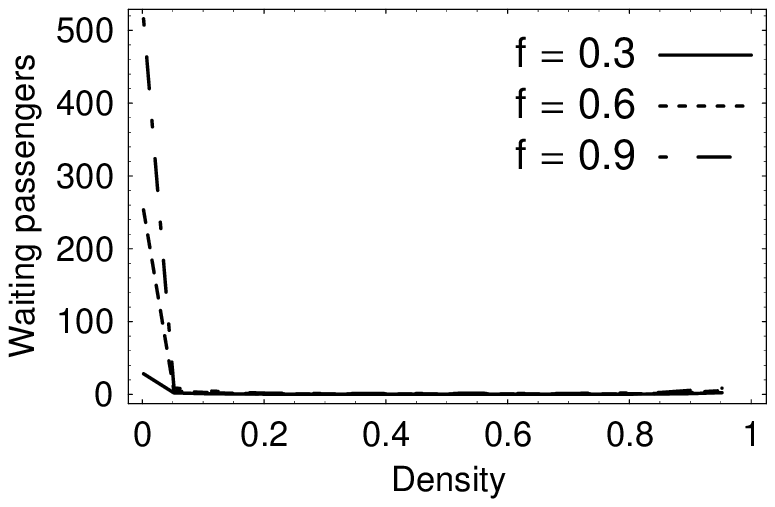}
\caption{The plot of $\langle V \rangle$ and $\langle N \rangle$ of the
 model B with information ($S=5$ and $f=0.3$, 0.6 and 0.9)}
\label{S=5_Ruletrue_info}
\end{center}
\end{figure}

\begin{figure}[h]
\begin{center}
\includegraphics[scale=0.75]{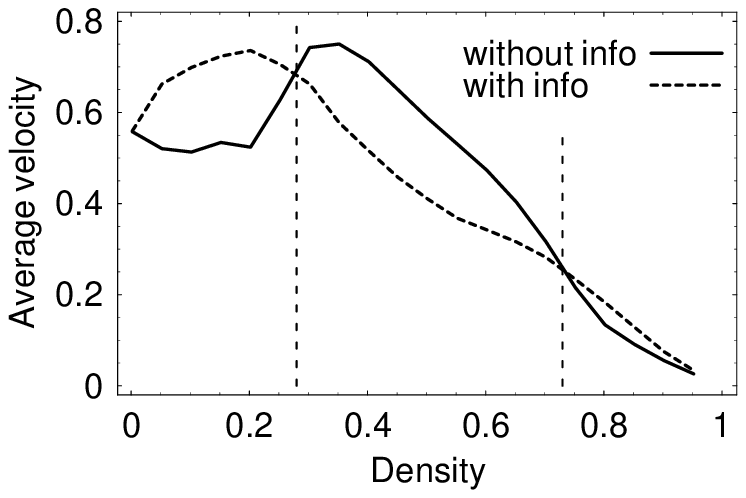}
\includegraphics[scale=0.75]{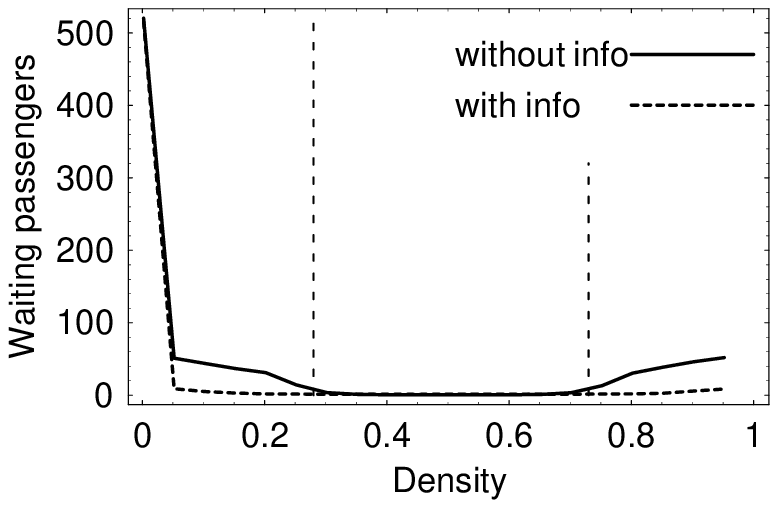}
\caption{The model B  with $S=5$ and $f=0.9$. 
The left vertical dash line is $\rho=0.28$ and the right is $\rho=0.73$
 in the two figures.}
\label{S=5_info}
\end{center}
\end{figure}

\subsection{PCM with information-based traffic control}

The results for the PCM with information-based traffic control system is 
shown in FIG.~$\ref{S=5_Ruletrue_info}$ and FIG.~$\ref{S=5_info}$.
In the FIG.~$\ref{S=5_Ruletrue_info}$ we plot $\langle V \rangle$ and
$\langle N \rangle$ against the density of buses for the parameter
$S=5$.  The density corresponding to the peak of the average velocity 
shifts to lower values when the information-based traffic control 
system is switched on. 

The data shown in FIG.~$\ref{S=5_info}$ establish that implementation 
of the information-based traffic control system does not necessarily 
always improve the efficiency of the public conveyance system. In 
fact, in the region $0.3 < \rho < 0.7$, the average velocity of the 
buses is higher if the information-based control system is switched 
off. Comparing $\langle V \rangle$ and $\langle N \rangle$ in
FIG.~\ref{S=5_info}, we find that information-based traffic control 
system can improves the efficiency by reducing the crowd of waiting 
passengers. But, in the absence of waiting passengers, introduction 
of the information-based control system adversely affects the 
efficiency of the public conveyance system by holding up the buses 
at bus stops when the number of buses in the next segment of the 
route exceeds $I_0$.

\begin{figure}[h]
\begin{center}
\includegraphics[scale=0.75]{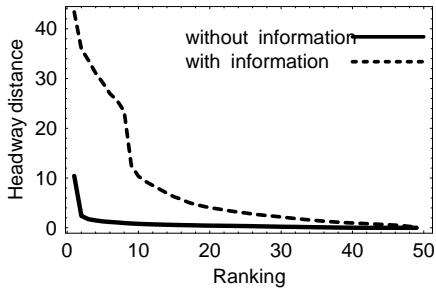}
\caption{Distribution of headway distance for
 $S=10$, $m=50$, $f=0.9$ in model B. This figure shows the plot of headway
 distance against the ranking.}
\label{headrank}
\end{center}
\end{figure}

Finally, FIG.~\ref{headrank} shows the distribution of headway distance
against the ranking, where we arrange the order of magnitude according
to the headway distance of buses in descending order. From this figure 
it is found that the headway distribution is dispersed by the effect of 
the information. The average headway distance with the information-based 
traffic control is equal to $8.34$, in contrast to a much shorter value 
of $0.66$ when that control system is switched off. Thus we confirm that 
the availability of the information $I_j$ and implementation of the 
traffic control system based on this information, significantly reduces 
the undesirable clustering of buses.

\section{MEAN FIELD ANALYSIS}

Let us estimate $\langle V \rangle$ theoretically in the low density 
limit $\rho \to 0$. Suppose, $T$ is the average time taken by a bus 
to complete one circuit of the route. In the model A, the number of 
hops made by a bus with probability $q$ during the time $T$ is $S$, 
i.e. the total number of bus stops. Therefore the average period $T$ 
for a bus in the model A is well approximated by
\begin{equation}
 T = \frac{L-S}{Q}+\frac{S}{q}
\label{time}
\end{equation}
and hence, 
\begin{equation}
 \langle V \rangle=\frac{L}{T}=\frac{LQq}{q(L-S)+QS}\,.
\label{av}
\end{equation}

In model B, in the  low density limit where $m$ buses
run practically unhindered and are distributed uniformly in the system 
without correlations, the average number of passengers $N$ waiting 
at a bus stop, just before the arrival of the next bus, is 
\begin{equation}
 N =\frac{f}{S}\left(\frac{\frac{L}{S}-1}{Q}+\frac{1}{q}\right)\frac{S}{m}. 
\label{N}
\end{equation}
The first factor $f/S$ on the right hand side of the equation (\ref{N}) 
is the probability of arrival of passengers per unit time. The second 
factor on the right hand side of (\ref{N}) is an estimate of the average 
time taken by a bus to traverse one segment of the route, i.e. the part 
of the route between successive bus stops. The last factor in the same 
equation is the average number of segments of the route in between two 
successive buses on the same route. Instead of the constant $q$ used in 
(\ref{av}) for the evaluation of $\langle V \rangle$ in the model A, we 
use 
\begin{equation}
\bar q =\frac{Q}{N+1}
\label{barq} 
\end{equation}
in eq.~(\ref{av}) and eq.~($\ref{N}$) for the model B. Then, for the model 
B, the hopping probability $Q$ is estimated self-consistently solving  
\begin{equation}
 \langle V \rangle = Q - \frac{f}{m}, 
\label{waitH}
\end{equation}
(\ref{av}) and (\ref{barq}) simultaneously.

We also obtain, for the model B, the average number of passengers
$\langle N \rangle$ waiting at a bus stop in the $\rho \to 0$ limit.
The average time for moving from one bus stop to the next is 
$\Delta t=(L/S-1)/Q+1/{\bar q}$ and, therefore, we have
\begin{eqnarray}
 \langle N \rangle &=& (f/S)\cdot(\Delta t + 2 \Delta t + \cdots + (S-1)\Delta
  t)/S\nonumber\\
&=&\frac{f(S-1)({\bar q}(L-S)+SQ)}{2S^2Q{\bar q}}.
\label{an}
\end{eqnarray}

As long as the number of waiting passengers does not exceed $N_{\rm max}$, 
we have observed reasonably good agreement between the analytical estimates 
(\ref{av}), (\ref{an}) and the corresponding numerical data obtained from 
computer simulations. For example, in the model A, we get the estimates 
$\langle V \rangle=0.85$ and $\langle N \rangle=1.71$ from the approximate 
mean field theory for the parameter set $S=50$, $m=1$, $Q=0.9$, $q=0.5$, 
$f=0.3$. The corresponding numbers obtained from direct computer 
simulations of the model A version of PCM are 0.84 and 1.78, respectively.
Similarly, in the model B under the same conditions, we get $\langle V
\rangle=0.60$ and $\langle N \rangle=2.45$ from the mean field theory,
while the corresponding numerical values are 0.60 and 2.51, respectively.
If we take sufficiently small $f$'s, then the mean-field estimates agree 
almost perfectly with the corresponding simulation data. However, our mean 
field analysis breaks down when a bus can not pick up all the passengers 
waiting at a bus stop.

\section{CONCLUDING DISCUSSIONS}

In this paper, we have proposed a public conveyance model (PCM) by using
stochastic CA. In our PCM, some realistic elements are introduced: e.g.,
the carrying capacity of a bus, 
the arbitrary number of bus stops, the halt time of a bus that depends 
on the number of waiting passengers, and an information-based bus traffic 
control system which reduces clustering of the buses on the given route. 

We have obtained quantitative results by using both computer simulations
and analytical calculations. In particular, we have introduced two 
different quantitative measures of the efficiency of the public conveyance 
system. We have found that the bus system works efficiently in a region 
of moderate number density of buses; too many or too few buses drastically 
reduce the efficiency of the bus-transport system. If the density of the 
buses is lower than optimal, not only large number of passengers are kept 
waiting at the stops for longer duration, but also the passengers in the 
buses get a slow ride as buses run slowly because they are slowed down 
at each stop to pick up the waiting passengers. On the other hand, if the
density of the buses is higher than optimal, the mutual hindrance created 
by the buses in the overcrowded route also lowers the efficiency of the 
transport system. Moreover, we have found that the average velocity 
increases, and the number of waiting passengers decreases, when the 
information-based bus traffic control system is switched on. However, 
this enhancement of efficiency of the conveyance system takes place 
only over a particular range of density; the information-based bus traffic 
control system does not necessarily improve the efficiency of the system 
in all possible situations.

We have compared two situations where the second situation is obtained 
from the first one by doubling the carrying capacity of each bus and 
reducing their number to half the original number on the same route. 
In the density region $\rho > 0.05$ the system of $N_{\rm max}=60$ is more 
efficient than that with $N_{\rm max}=120$. However, at small densities 
($\rho < 0.05$), although the average velocity increases, the number of 
waiting passengers also increases, by doubling the carrying capacity 
from $N_{\rm max}=60$ to $N_{\rm max} = 120$. Hence, bus-transport system 
administrators would face a dilemma in this region of small density.

Finally, in our PCM, the effect of the disembarking passengers on the 
halt time of the buses has not been captured explicitly. Moreover,
this study is restricted to periodic boundary conditions. The clustering
of particles occurs not only in a ring-like bus route, but also in 
shuttle services of buses and trains. Thus it would be interesting to 
investigate the effects of the information-based traffic control system 
also on such public transport systems. In a future work, we intend to 
report the results of our investigations of the model under non-periodic  
boundary conditions.
We hope our model will help in understanding the 
mechanism of congestion in public conveyance system and will provide 
insight as to the possible ways to reduce undesirable clustering of the 
vehicles.

\vspace{0.5cm}

\noindent{\bf Acknowledgments}: Work of one of the authors (DC) has been 
supported, in part, by the Council of Scientific and Industrial Research 
(CSIR), government of India.

\end{document}